\documentclass[%
 aip,
 jmp,%
 amsmath,amssymb,
 reprint,%
]{revtex4-2}

\usepackage{graphicx}
\usepackage{dcolumn}
\usepackage{bm}
\usepackage{comment}
\usepackage{xcolor}

\begin{document}

\preprint{AIP/123-QED}

\title{Pressure-Strain Interaction: III. Particle-in-Cell Simulations of Magnetic Reconnection}

\author{M. Hasan Barbhuiya}
\email{mhb0004@mix.wvu.edu}
\author{Paul A. Cassak}
\affiliation{Department of Physics and Astronomy and the Center for KINETIC Plasma Physics, \\
West Virginia University, Morgantown, WV 26506, USA
}

\date{\today}

\begin{abstract}
How energy is converted into thermal energy in weakly collisional and collisionless plasma processes such as magnetic reconnection and plasma turbulence has recently been the subject of intense scrutiny.  The pressure-strain interaction has emerged as an important piece, as it describes the rate of conversion between bulk flow and thermal energy density. In two companion studies, we presented an alternate decomposition of the pressure-strain interaction to isolate the effects of converging/diverging flow and flow shear instead of compressible and incompressible flow, and we derived the pressure-strain interaction in magnetic field-aligned coordinates. Here, we use these results to study pressure-strain interaction
during two-dimensional anti-parallel magnetic reconnection.  We perform particle-in-cell simulations and plot the decompositions in both Cartesian and magnetic field-aligned coordinates. We identify the mechanisms contributing to positive and negative pressure-strain interaction during reconnection. This study provides a roadmap for interpreting numerical and observational data of the pressure-strain interaction, which should be important for studies of reconnection, turbulence, and collisionless shocks.
\end{abstract}


\maketitle

\section{Introduction}
\label{sec:intro}

The pressure-strain interaction has garnered significant attention in the past few years because it describes the rate of conversion between bulk flow and thermal energy density [see Ref.~\cite{Cassak_PiD1_2022} and references therein].  The pressure-strain interaction is written equivalently as 
\begin{equation}
    -({\bf P} \cdot \boldsymbol{\nabla}) \cdot {\bf u} = -{\bf P} : {\bf S},
\end{equation}
where ${\bf P}$ is the pressure tensor, ${\bf u}$ is the bulk flow velocity, ${\bf S} = \boldsymbol{\nabla} {\bf u}$ is the strain rate tensor, and the minus sign is included so that a positive value denotes a contribution to increasing the thermal energy density. The strain rate tensor can be decomposed into the bulk flow divergence $(1/3){\bf I} (\boldsymbol{\nabla} \cdot {\bf u})$ describing compression and $\boldsymbol{{\cal D}}$ describing incompressible flow \cite{del_sarto_pressure_2018}, where ${\bf I}$ is the identity tensor and $\boldsymbol{{\cal D}}$ is the traceless strain rate tensor with elements ${\cal D}_{jk} = (1/2) (\partial u_{j} / \partial r_k + \partial u_{k} / \partial r_j) - (1/3) \delta_{jk} (\boldsymbol{\nabla} \cdot {\bf u})$, where $\delta_{jk}$ is the Kroenecker delta.  The pressure-strain interaction is then decomposed as \cite{del_sarto_pressure_2016,Yang17,yang_PRE_2017,del_sarto_pressure_2018}
\begin{equation}
  -({\bf P} \cdot \boldsymbol{\nabla}) \cdot {\bf u} = - {\cal P} (\boldsymbol{\nabla} \cdot {\bf u}) - \boldsymbol{\Pi} : \boldsymbol{{\cal D}}, \label{eq:pdelu}
\end{equation}
where ${\cal P} = (1/3) P_{jj}$ is the effective pressure and $\boldsymbol{\Pi} = {\bf P} - {\cal P} {\bf I}$ is the deviatoric pressure tensor.  The first and second terms in Eq.~(\ref{eq:pdelu}), including the minus signs, are the pressure dilatation and the term dubbed ${\rm Pi-D}$, respectively \cite{Yang17}. ${\rm Pi-D}$ is the collisionless analogue of the viscous heating rate \cite{yang_PRE_2017}.

In the first study of a three part series \cite{Cassak_PiD1_2022} (``Paper I''), we introduced an alternate decomposition to Eq.~(\ref{eq:pdelu}). Rather than isolating compressible and incompressible heating/cooling, it isolates the effect of flow convergence/divergence in a term we call ${\rm PDU}$ and flow shear in a term we call ${\rm Pi-D}_{{\rm shear}}$. Analytically, $-{\bf P} : {\bf S} = {\rm PDU}+{\rm Pi-D}_{{\rm shear}}$, where (in Cartesian coordinates)
\begin{subequations}
\begin{eqnarray}
    {\rm PDU} & = & -\left(P_{xx} \frac{\partial u_x}{\partial x} + P_{yy} \frac{\partial u_y}{\partial y} + P_{zz} \frac{\partial u_z}{\partial z}\right), \label{eq:pscompress} \\
    {\rm Pi-D}_{{\rm shear}} & = & -\left[P_{xy} \left( \frac{\partial u_x}{\partial y} + \frac{\partial u_y}{\partial x} \right) + P_{xz} \left( \frac{\partial u_x}{\partial z} + \frac{\partial u_z}{\partial x} \right) + P_{yz} \left( \frac{\partial u_y}{\partial z} + \frac{\partial u_z}{\partial y} \right) \right]. \label{eq:psincompress}
\end{eqnarray}
\end{subequations}
Equation~(\ref{eq:pscompress}) is a sum of pressure dilatation (compression/expansion) and normal deformation (the part of {\rm Pi-D} coming from the diagonal elements of $\boldsymbol{{\cal D}}$ describing an incompressible change of shape of a fluid element), while Eq.~(\ref{eq:psincompress}) is the part of ${\rm Pi-D}$ coming from the off-diagonal elements of $\boldsymbol{{\cal D}}$ related to bulk flow shear.  

In the second study in the series \cite{Cassak_PiD2_2022} (``Paper II''), we calculated the decomposition of the pressure-strain interaction in magnetic field-aligned coordinates, finding eight sets of terms:
\begin{subequations}
\begin{eqnarray}
-PS_1 & = & -P_{\|} (\nabla_\| u_\|), \\
-PS_2 & = & -P_{\kappa \kappa} (\nabla_\kappa u_\kappa) -P_{nn} (\nabla_n u_n), \\
-PS_3 & = & -P_{\kappa b} (\nabla_\kappa u_\|) - P_{nb} (\nabla_n u_\|), \\
-PS_4 & = & -P_{b\kappa} (\nabla_\| u_\kappa) - P_{bn} (\nabla_\| u_n) - P_{\kappa n} (\nabla_\kappa u_n) - P_{n \kappa} (\nabla_n u_\kappa), \\
-PS_5 & = & u_\kappa \left( P_\| + P_{b \kappa} W_\kappa + P_{b n} W_n \right) \kappa = u_\kappa P_{b \alpha} W_\alpha \kappa  \\
  -PS_6 & = & -u_\kappa \left(P_{bn} + P_{\kappa n} W_\kappa + P_{nn} W_n \right) \tau = -u_\kappa P_{n \alpha} W_\alpha \tau  \\
  -PS_7 & = & - u_\| \left( P_{b \kappa} + P_{\kappa \kappa} W_\kappa + P_{\kappa n} W_n \right) \kappa = -u_\| P_{\kappa \alpha} W_\alpha \kappa \\
  -PS_8 & = & u_n \left(P_{b \kappa} + P_{\kappa \kappa} W_\kappa + P_{n \kappa} W_n \right) \tau = u_n P_{\kappa \alpha} W_\alpha \tau,
\end{eqnarray}
\end{subequations}
where the $b$ and $\|$ subscripts denote the component parallel to the magnetic field ${\bf B}$, the $\kappa$ subscript denotes the component in the direction of the magnetic field line curvature $\boldsymbol{\kappa} = {\bf \hat{b}} \cdot \boldsymbol{\nabla} {\bf \hat{b}}$, where ${\bf \hat{b}} = {\bf B} / |{\bf B}|$, and the $n$ subscript denotes the component in the binormal direction  ${\bf \hat{n}} = {\bf \hat{b}} \times \boldsymbol{\hat{\kappa}}$, where $\boldsymbol{\hat{\kappa}} = \boldsymbol{\kappa}/|\boldsymbol{\kappa}|$.  The quantities $\kappa = |\boldsymbol{\kappa}| = |{\bf \hat{b}} \cdot \boldsymbol{\nabla} {\bf \hat{b}}|$ and $\tau = - \boldsymbol{\hat{\kappa}} \cdot \nabla_\| {\bf \hat{n}}$ are the magnetic field line curvature and torsion, respectively.  The quantity ${\bf W}$ is a vector with components $W_b = 1, W_\kappa = (\boldsymbol{\hat{\kappa}} \cdot \nabla_\kappa {\bf \hat{b}})/\kappa$, and $W_n = -(\boldsymbol{\hat{\kappa}} \cdot \nabla_n {\bf \hat{n}})/\tau$. These components are the prefactors linking gradients in the curvature and binormal directions to the parallel direction such that $\nabla_\kappa =  W_\kappa \nabla_\|$, and $\nabla_n  =  W_n \nabla_\|$.  It was shown in Ref.~\cite{Cassak_PiD2_2022} that these eight terms correspond physically to parallel flow compression ($-PS_1$), perpendicular flow compression ($-PS_2$), shear of parallel flow in the perpendicular direction ($-PS_3$), shear of perpendicular flow in the parallel and/or perpendicular directions ($-PS_4$), perpendicular geometrical compression ($-PS_5$), torsional geometrical compression ($-PS_6$), parallel geometrical shear ($-PS_7$), and torsional geometrical shear ($-PS_8$). Here and in what follows, we simplify the wording by using compression to mean positive (compression) or negative (expansion) effects.

In this study, we calculate the terms in the decomposition of the pressure-strain interaction in Cartesian and magnetic field-aligned coordinates in two-dimensional particle-in-cell (PIC) simulations of anti-parallel symmetric magnetic reconnection. The purposes for this study are two-fold.  First, we demonstrate a roadmap for using the analytical results in Papers I and II to study the pressure-strain interaction in weakly collisional or collisionless plasmas, using magnetic reconnection as an example.  Second, understanding the rate of conversion between bulk flow and thermal energy density during reconnection is of intrinsic interest, and has been the subject of numerical and observational studies \cite{Sitnov18,Chasapis18,bandyopadhyay_energy_2021,Pezzi21}.  An outcome of the present study is a map of an electron diffusion region identifying where energy conversion via pressure-strain interaction occurs and the physical causes of it in each location.  Since ${\rm Pi-D}$ has been a significant topic of research, including the realization that it can be negative \cite{Yang17,zhou_measurements_2021}, we also identify its physical cause during reconnection.

The layout of this manuscript is as follows.  Section~\ref{sec:sims} gives the details of the numerical simulation setup. Sections~\ref{sec:simresults} and \ref{sec:simresultsfieldaligned} give the numerical results in Cartesian and magnetic field-aligned coordinates, respectively. Section~\ref{sec:discuss} includes a discussion and conclusions.

\section{Numerical Simulation Setup}
\label{sec:sims}

To calculate the contributions to the pressure-strain interaction in a numerical simulation, we use the particle-in-cell code {\tt p3d} \cite{zeiler:2002} to simulate symmetric anti-parallel magnetic reconnection. The simulations are 2.5D in position space and 3D in velocity-space. The particles are stepped forward in time using the relativistic Boris particle stepper \cite{birdsall91a}, while electromagnetic fields are stepped forward using the trapezoidal leapfrog \cite{guzdar93a}. The time step for the fields is half the time step for the particles. The multigrid method \cite{Trottenberg00} is used to clean the electric field ${\bf E}$ every 10 particle time-steps to enforce Poisson's equation.  The boundary conditions are periodic in both spatial directions.  

Quantities produced by the simulation are in normalized units. The initial asymptotic reconnecting magnetic field strength is $B_0$, and $n_0$ is the plasma number density at the center of the current sheet minus the ambient background plasma density.  Lengths, velocities, times, and temperatures are normalized to the ion inertial scale $d_{i0} = c/\omega_{pi0}$, the  Alfv\'en speed $c_{A0} = B_0/(4 \pi m_i n_0)^{1/2}$, the inverse ion cyclotron frequency $\Omega_{ci0}^{-1}= (e B_{0} / m_{i} c)^{-1}$, and $m_i c_{A0}^2/k_B$, respectively, where $\omega_{pi0} = (4 \pi n_0 e^2 /m_i)^{1/2}$ is the ion plasma frequency, $e$ is the ion charge, $m_i$ is the ion mass, and $c$ is the speed of light.  Consequently, power densities making up the pressure-strain interaction are in units of $\Omega_{ci0} (B_0^2/4 \pi)$. 

The speed of light is $c=15$, which is sufficient for the purposes of the present study.  The electron to ion mass ratio is $m_e/m_i = 0.04$, which is relatively high.  We expect this choice could influence the amplitude and spatial size of structures in the electron diffusion region \cite{Ricci02,Guo07}, but we do not expect it to affect the qualitative structure \cite{Du18}. The scaling of the results with electron mass is discussed further in Secs.~\ref{sec:contributions} and \ref{sec:discuss}. The simulation domain size is $L_x \times L_y = 12.8 \times 6.4$. We use $1024 \times 512$ grid cells, and initially use 25,600 weighted particles per grid. The grid scale is $\Delta=0.0125$, smaller than the smallest length-scale of the system, the electron Debye length $\lambda_{De}=0.0176$. The time-step is $\Delta t=0.001$, smaller than the smallest time-scale of the system, the inverse electron plasma frequency $\omega^{-1}_{pe}=0.012$.

The initial conditions are a standard double tanh magnetic field.  The initial magnetic field profile is
\begin{equation}
B_x(y) = \tanh{\left(\frac{y-L_y/4}{w_0}\right)} - \tanh{\left(\frac{y-3L_y/4}{w_0}\right)} - 1,
\end{equation}
where $w_0=0.5$ is the initial half-thickness of the current sheet.  There is no initial out-of-plane (guide) magnetic field.  The electrons and ions are initially drifting Maxwellian distributions with density profiles given by
\begin{equation}
n(y) = \frac{1}{2(T_{e}+T_{i})} \left[ {\rm sech}^{2}\left(\frac{y-L_y/4}{w_0}\right) + {\rm sech}^{2}\left(\frac{y-3L_y/4}{w_0}\right) \right] + n_{up},
\end{equation}
where $n_{up}=0.2$ is the initial plasma density far from the current sheet. The temperature of the background plasma is initially uniform, with electron temperature $T_{e}=1/12$ and ion temperature $T_{i}=5/12$. A magnetic perturbation $\tilde{{\bf B}} = - {\bf \hat{z}} \times \nabla \tilde{\psi}$ is used to initiate reconnection, where 
\begin{equation}
\tilde{\psi} = -\frac{\tilde{B}L_y}{4\pi} \sin \left(\frac{2 \pi x}{L_x} \right) \left[1-\cos\left(\frac{4\pi y}{L_y}\right)\right],
\end{equation}
and the perturbation amplitude is $\tilde{B} = 0.05$.

All simulation data are shown from the lower current sheet at $t = 13$, when the reconnection rate is most rapidly increasing from zero to its maximum value, {\it i.e.,} it is not in the steady state.  To reduce PIC noise, we recursively smooth the raw simulation quantities four times over a width of four cells, then we take the necessary spatial derivatives, and finally the results are recursively smoothed four times over four cells again. This level of smoothing is decided on by trying a number of different options for the number of cells and how many recursions, while confirming that smoothing does not greatly alter the signal structure. We focus on the electrons for this study; in what follows, the $e$ subscript denoting electrons is suppressed for simplicity except where needed for clarity.

\section{Simulation results - Cartesian Coordinates}
\label{sec:simresults}

\subsection{Overview}
\label{sec:overview1}

\begin{figure}
\includegraphics[width=5.4in]{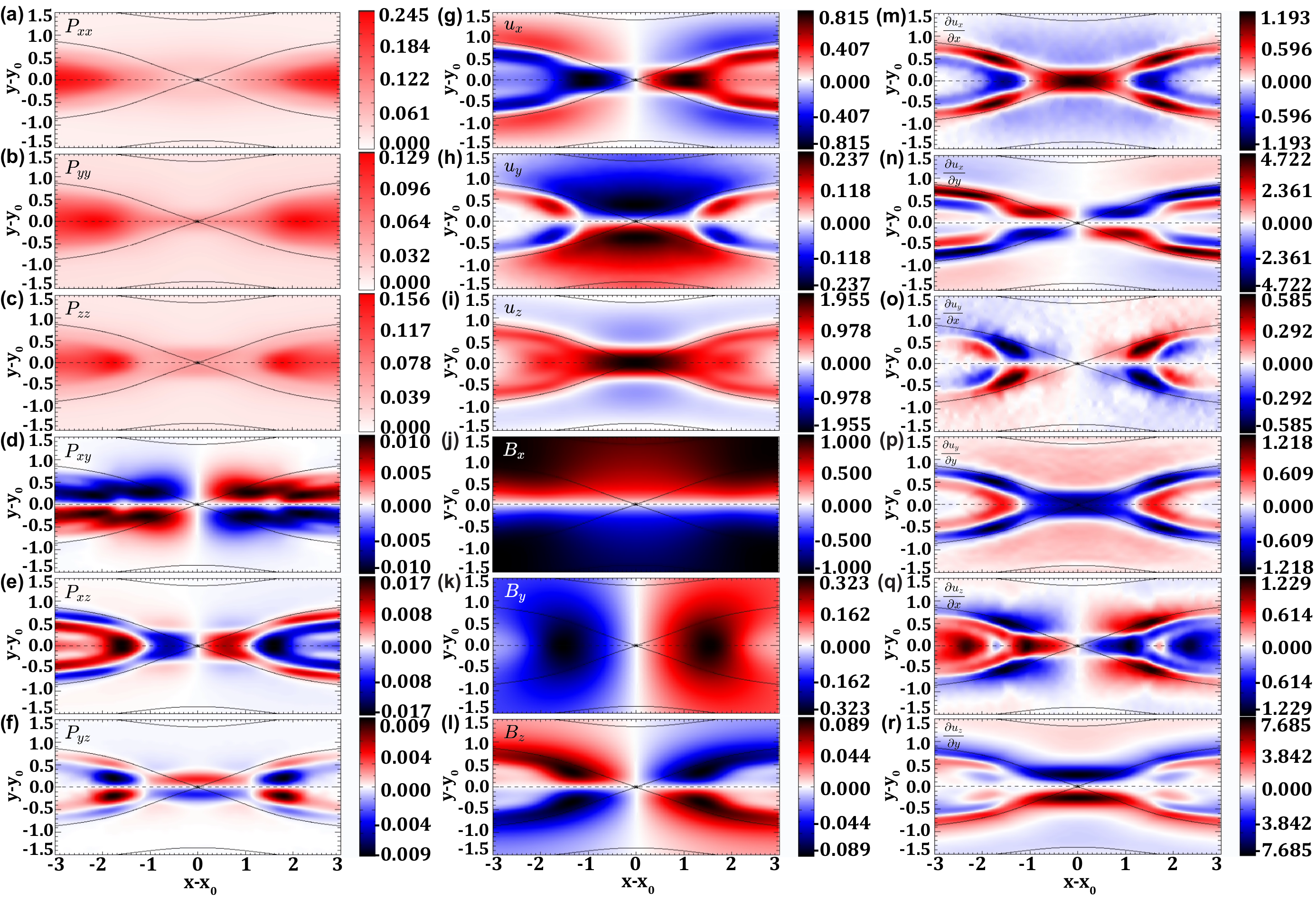}
\caption{\label{fig:2dprofiles} Two-dimensional profiles of the (a)-(f) six independent elements of the electron pressure tensor ${\bf P}$, (g)-(i) three components of bulk electron velocity ${\bf u}$, (j)-(l) three components of the magnetic field ${\bf B}$, and (m)-(r) six non-zero elements of the strain-rate tensor $\boldsymbol{\nabla} {\bf u}$, as labeled in each panel. Representative magnetic field projections in the $xy$ plane are in black.}
\end{figure}

The 2D profiles of the quantities that go into the calculation of the pressure-strain interaction in Cartesian coordinates and the magnetic field are provided in Fig.~\ref{fig:2dprofiles}. It contains the (a)-(f) six independent elements of the electron pressure tensor ${\bf P}$, (g)-(i) three components of bulk electron velocity ${\bf u}$, (j)-(l) three components of the magnetic field ${\bf B}$, and (m)-(r) six non-zero elements of strain rate tensor $\boldsymbol{\nabla} {\bf u}$ (since $\partial/\partial z = 0$ for this 2D simulation). Representative magnetic field projections in the $xy$ plane are included in black for perspective. The plots represent only a portion of the computational domain centered at the X-line location $(x_0,y_0)$ from $|x - x_0| < 3$ and $|y - y_0| < 1.5$.  This encompasses the electron diffusion region (EDR) which is approximately $|x-x_0| < 2,|y-y_0| < 0.35$. 

\subsection{Decomposition of Pressure-Strain Interaction in Cartesian Coordinates}
\label{sec:cartesiansims1}

\begin{figure}
\includegraphics[width=5.4in]{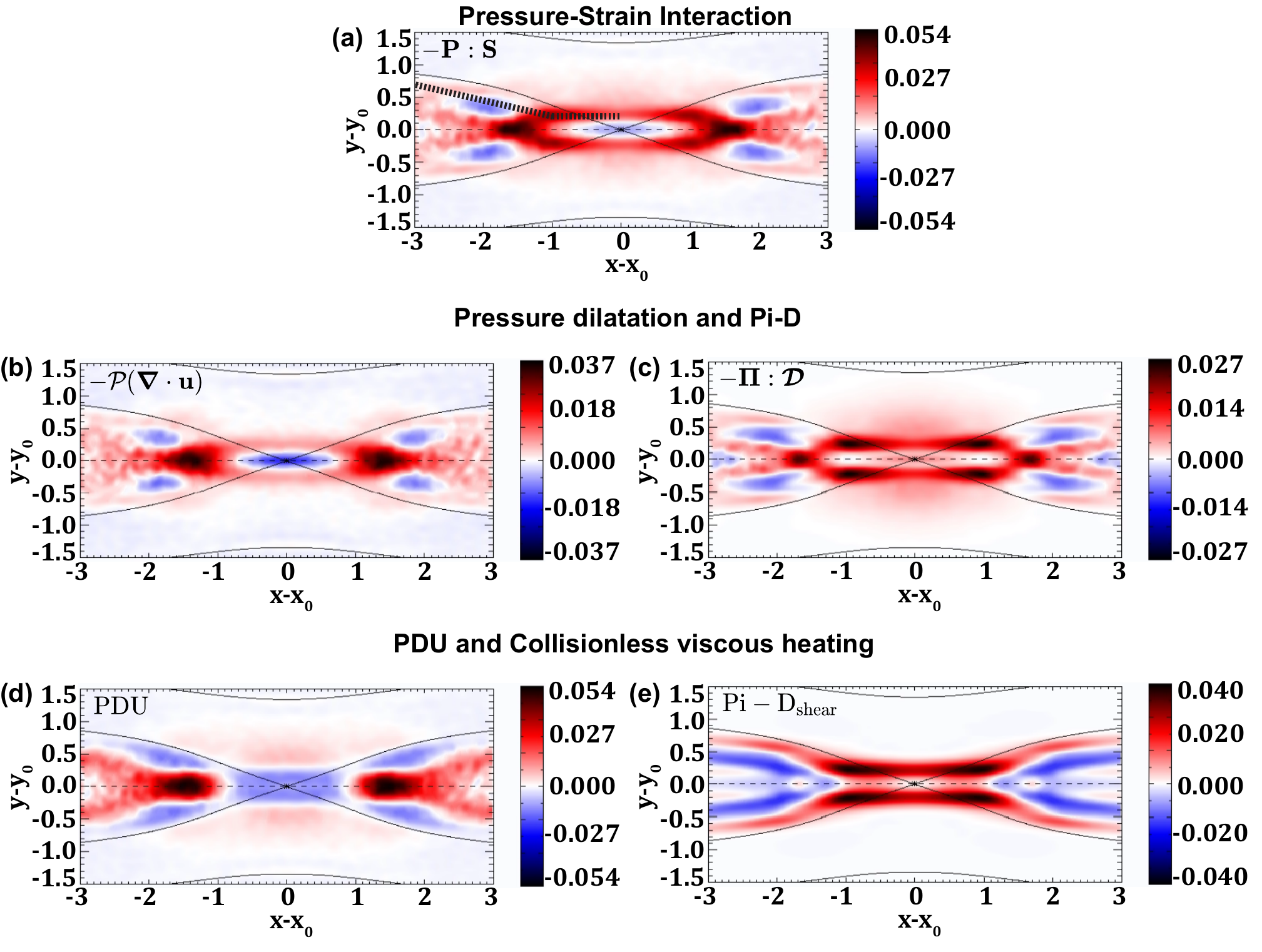}
\caption{\label{fig:2TermDecomp} Pressure-strain interaction for electrons in a reconnection simulation. (a) The pressure-strain interaction $-{\bf P}:{\bf S}$. (b) Pressure dilatation $-{\cal P} (\boldsymbol{\nabla} \cdot {\bf u})$ and (c) ${\rm Pi-D}$, giving the compressible and incompressible parts. (d) ${\rm PDU}$ and (e) ${\rm Pi-D}_{{\rm shear}}$, giving the flow converging/diverging and flow shear parts. The dotted-line in (a) is the path along which 1D cuts are taken in Figs.~\ref{fig:1dcuts} and \ref{fig:9TermDecomp1Dcuts}.}
\end{figure}

The pressure-strain interaction and two decompositions in Cartesian coordinates are shown in Fig.~\ref{fig:2TermDecomp}. The color bars have red values for positive and blue for negative, and their ranges are different for the different plots. In this subsection, we largely focus on a qualitative comparison of the pressure-strain interaction and the decompositions in question.  In the subsequent one, we extract the physical causes.  The pressure-strain interaction $-{\bf P} : {\bf S}$ is shown in Fig.~\ref{fig:2TermDecomp}(a). The pressure-strain interaction at the X-line is negative, {\it i.e.,} in isolation it would lead to a decrease in thermal energy. We reiterate that although the thermal energy flux and heat flux must integrate to zero over the whole periodic domain, they are not necessarily zero locally. Thus, one cannot conclude that there is a net decrease in thermal energy at the X-line simply because $-{\bf P}:{\bf S}$ is negative, only that the pressure-strain interaction by itself would lead to a decrease in thermal energy.  (Indeed, in a separate study \cite{Cassak_FirstLaw_2022}, we show that the thermal energy at the X-line is increasing at the time shown.) The edges of the EDR indicate a net positive pressure-strain interaction.
The downstream region $|x-x_0| \simeq 2$ reveals a positive pressure-strain interaction near the neutral line $y = y_0$, immediately surrounded by a region of negative pressure-strain interaction out to about $|y-y_0| \simeq 0.5$.  

The pressure dilatation $-{\cal P} (\boldsymbol{\nabla} \cdot {\bf u})$ and ${\rm Pi-D}$ are shown in Figs.~\ref{fig:2TermDecomp}(b) and (c), respectively. The pressure dilatation gives the contribution due to compression/expansion, while ${\rm Pi-D}$ is due to incompressible effects. They have previously been plotted in 2D simulations of reconnection \cite{Pezzi21} (see their Fig.~2).  
We believe the present data looks sharper because it is zoomed in closer, we employ more particles per grid and smoothing, and our data is plotted during the phase when the reconnection rate is increasing most rapidly rather than after the reconnection rate reaches its maximum value.  We see a coherent region of negative $-{\cal P} (\boldsymbol{\nabla} \cdot {\bf u})$ and ${\rm Pi-D}$ near the downstream edge of the EDR at $|x-x_0| \simeq 2, |y-y_0| \simeq 0.35$, not seen in Ref.~\cite{Pezzi21}, as will be discussed further in the next subsection.

The decomposition of $-{\bf P}:{\bf S}$ into ${\rm PDU}$ and ${\rm Pi-D}_{{\rm shear}}$, defined in Eqs.~(\ref{eq:pscompress}) and (\ref{eq:psincompress}), respectively, are plotted in Figs.~\ref{fig:2TermDecomp}(d) and (e). While the pressure dilatation and ${\rm Pi-D}$ each have similar overall structure to $-{\bf P}:{\bf S}$, we find ${\rm PDU}$ and ${\rm Pi-D}_{{\rm shear}}$ have qualitative dissimilarities compared to the structure of $-{\bf P}:{\bf S}$. In ${\rm PDU}$, the entire inner part of the EDR is negative, as opposed to only a small region near the X-line in $-{\bf P}:{\bf S}$.  Perhaps most importantly, the region of highest pressure-strain interaction, the downstream edge of the EDR, shows up entirely in ${\rm PDU}$, and it is essentially zero in ${\rm Pi-D}_{{\rm shear}}$.  This shows the contribution to positive pressure-strain interaction is due entirely to the converging flows at the edge of the EDR.  The decomposition given by pressure-dilatation and ${\rm Pi-D}$ reveal that the converging flow is associated with both compression and incompressible deformation.  A similar example is in the upstream region, where there is weak positive pressure-strain interaction.  Similar structure is seen in ${\rm PDU}$, while ${\rm Pi-D}_{{\rm shear}}$ is nearly zero everywhere upstream of the EDR.  Since ${\rm Pi-D}$ is non-zero upstream of the EDR, we can conclude that incompressible deformation is the cause of this contribution to the pressure-strain interaction. We similarly note that there is a relatively strong positive pressure-strain interaction at the upstream edges of the EDR (the horizontal red bands), and it is caused nearly completely by flow shear.  The cause of this positive pressure-strain interaction is more ambiguous in the pressure dilatation and ${\rm Pi-D}$ decomposition, where both compressible and incompressible effects contribute.  These three examples imply that ${\rm PDU}$ and ${\rm Pi-D}_{{\rm shear}}$ can be useful to help separate out the key physical cause of the pressure-strain interaction in these three regions of interest. In summary, both decompositions provide useful information about the causes of the pressure-strain interaction, and using the different decompositions together can help identify the key physical causes of the pressure-strain interaction.

\subsection{Largest Contributions to Pressure-Strain Interaction in Cartesian Coordinates}
\label{sec:contributions1}

Here, we discuss the regions of most significant contributions to the pressure-strain interaction and use the decompositions to understand their physical causes.  The region of the highest contribution to the pressure-strain interaction is the downstream edge of the EDR, $1 < |x-x_0| < 2$ and $|y-y_0| < 0.3$, as seen in Fig.~\ref{fig:2TermDecomp}(a). As discussed briefly in the previous subsection, the cause of this is the converging flow when the electron jet from the EDR impacts the magnetic island [see Figs.~\ref{fig:2dprofiles}(a) and (m) and Fig.~\ref{fig:2TermDecomp}(d)].  Both compression and (incompressible) normal deformation are taking place in this region, which is why both pressure-dilatation and ${\rm Pi-D}$ are non-zero in this region [Fig.~\ref{fig:2TermDecomp}(b) and (c)].

In the ion diffusion region (IDR) significantly upstream of the EDR $|x-x_0| < 1, 0.35 < |y-y_0| < 2.24$, electrons decouple from the ions at the upstream edge of the IDR and then accelerate towards the X-line due to the Hall electric field. This leads to expansion, associated with cooling.  This shows up as the weakly blue region in $-{\cal P} (\boldsymbol{\nabla} \cdot {\bf u})$ and ${\rm PDU}$ outside $|y-y_0| \simeq 0.8$, with the inflow gradient profile in Fig.~\ref{fig:2dprofiles}(p).  Then, the electrons slow down upon reaching the upstream edge of the EDR, $|y-y_0| <0.35$ [Fig.~\ref{fig:2dprofiles}(p)], which causes compression and is associated with heating.  This shows up in ${\rm PDU}$ but not ${\rm Pi-D}_{{\rm shear}}$, meaning both compression and normal deformation are taking place.  The normal deformation describes the change of shape of the phase space density, which are known to elongate in the parallel direction due to electron trapping \cite{Egedal13}.  This is to be contrasted with the decomposition in terms of the pressure dilatation and ${\rm Pi-D}$, where this effect shows up in ${\rm Pi-D}$ because normal deformation is one of the two terms within ${\rm Pi-D}$.

Surrounding the X-line at $|x-x_0| < 0.7, |y-y_0| < 0.35$, there is a region of negative pressure-strain interaction [Fig.~\ref{fig:2TermDecomp}(a)]. This is caused by the acceleration of electrons into the exhaust jet; the $u_x$ flow increases in magnitude away from the X-line [Fig.~\ref{fig:2dprofiles}(m)], which is an expansion of the plasma. This shows up as the negative region near the origin in $-{\cal P}(\boldsymbol{\nabla} \cdot {\bf u})$ and ${\rm PDU}$. Since ${\rm Pi-D}$ is small near the X-line, we immediately conclude that compression is 
the most important effect here.

There are two other regions of significant pressure-strain interaction [Fig.~\ref{fig:2TermDecomp}(a)], namely the positive region at the upstream edge of the EDR ($|x-x_0|<1.5, 0.2 < |y-y_0| < 0.35$), and the negative region at the downstream edge of the EDR just inside the separatrices ($1 < |x-x_0| < 2, |y-y_0| \simeq 0.5$).  For the positive region at the upstream edge of the EDR, ${\rm Pi-D}_{{\rm shear}}$ provides the physical cause.  As there is no comparable signal in ${\rm PDU}$, we conclude it is caused solely by flow shear.  The plots of $-{\cal P}(\boldsymbol{\nabla} \cdot {\bf u})$ and ${\rm Pi-D}$ both show signals in this region, which implies that both compression and normal deformation are playing a role but are actually nearly canceling out. Two effects lead to the bulk flow shear that leads to this positive pressure-strain interaction.  First, the rapid drop-off of the out-of-plane flow $u_{z}$ in the inflow direction [Fig.~\ref{fig:2dprofiles}(i)] gives rise to a significant bipolar $\partial u_{z}/\partial y$ [Fig.~\ref{fig:2dprofiles}(r)].  This is in the same location as a bipolar pressure anisotropy $P_{yz}$ [Fig.~\ref{fig:2dprofiles}(f)], which conspires with the flow shear to give a positive ${\rm Pi-D}_{{\rm shear}}$ in the region $|x-x_0| < 1$.  In addition, the outflow $u_{x}$ [Fig.~\ref{fig:2dprofiles}(g)] rapidly changes in the inflow direction, leading to a quadrupolar $\partial u_{x} / \partial y$ [Fig.~\ref{fig:2dprofiles}(n)]. There is a quadrupolar $P_{xy}$ [Fig.~\ref{fig:2dprofiles}(d)], which conspires with the flow to give a positive ${\rm Pi-D}_{{\rm shear}}$ in the region $0.5 < |x-x_0| < 1.5$.  

\begin{figure}
\includegraphics[width=3.4in]{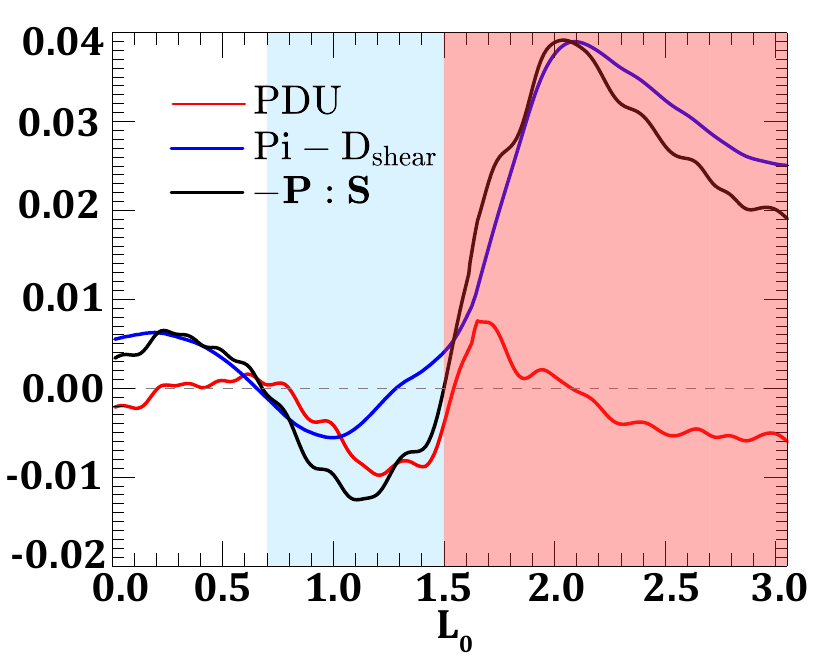}
\caption{\label{fig:1dcuts} Pressure-strain interaction along the 1D path shown in Fig.~\ref{fig:2TermDecomp}(a).  $L_0$ is the distance along the dotted path from the left.  $-{\bf P}:{\bf S}$ is in black, ${\rm PDU}$ is in red, and ${\rm Pi-D}_{{\rm shear}}$ is in blue. The shading marks the regions of negative (blue) and positive (red) pressure-strain interaction}.
\end{figure}

To see the contributions more clearly, we plot the profiles of the pressure-strain interaction $-{\bf P} : {\bf S}$ (black), ${\rm PDU}$ (red), and ${\rm Pi-D}_{{\rm shear}}$ (blue) in Fig.~\ref{fig:1dcuts} along the cut displayed as the black dotted path in Fig.~\ref{fig:2TermDecomp}(a). This path goes through the region of negative pressure-strain interaction in the exhaust and positive pressure-strain interaction along the upstream edge of the EDR.  The distance along the path starting from the left is $L_0$. The plot shows that the region of positive pressure-strain interaction at the upstream edge of the EDR, shaded in red, is due to ${\rm Pi-D}_{{\rm shear}}$, as inferred from the 2D plots in Fig.~\ref{fig:2TermDecomp}.  

Finally, for the region of negative pressure-strain interaction shaded in blue, closer to the X-line the dominant contribution is diverging flow (${\rm PDU}$), which occurs as the electron exhaust gets deflected around the island and accelerates away from the neutral line.  Further from the X-line, bulk flow shear due to the localized electron beam going around the island becomes equally important.  This region of negative pressure-strain interaction is not seen in the simulations in Ref.~\cite{Pezzi21}.  It is possibly due to the expansion caused by the separatrix opening out in time as reconnection onsets in our simulations, which would not have been seen in Ref.~\cite{Pezzi21} because their data were from a time after the maximum in the reconnection rate. Further work would be necessary to confirm or refute this possibility.

Since the negativity of ${\rm Pi-D}$ has been an important topic of consideration in the recent literature, we also discuss it here. By taking cuts of the deformation and shear parts of ${\rm Pi-D}$ along the same black dotted path in Fig.~\ref{fig:2TermDecomp}(a) (not shown), we find that the region of negative ${\rm Pi-D}$ in the downstream region (centered around $|x-x_0| \approx 2, |y-y_0| \approx 0.3$) is due to flow shear rather than normal deformation.

\section{Simulation Results - Magnetic Field-Aligned Coordinates}
\label{sec:simresultsfieldaligned}

\subsection{Overview}
\label{sec:overview}

\begin{figure}
\includegraphics[width=5.4in]{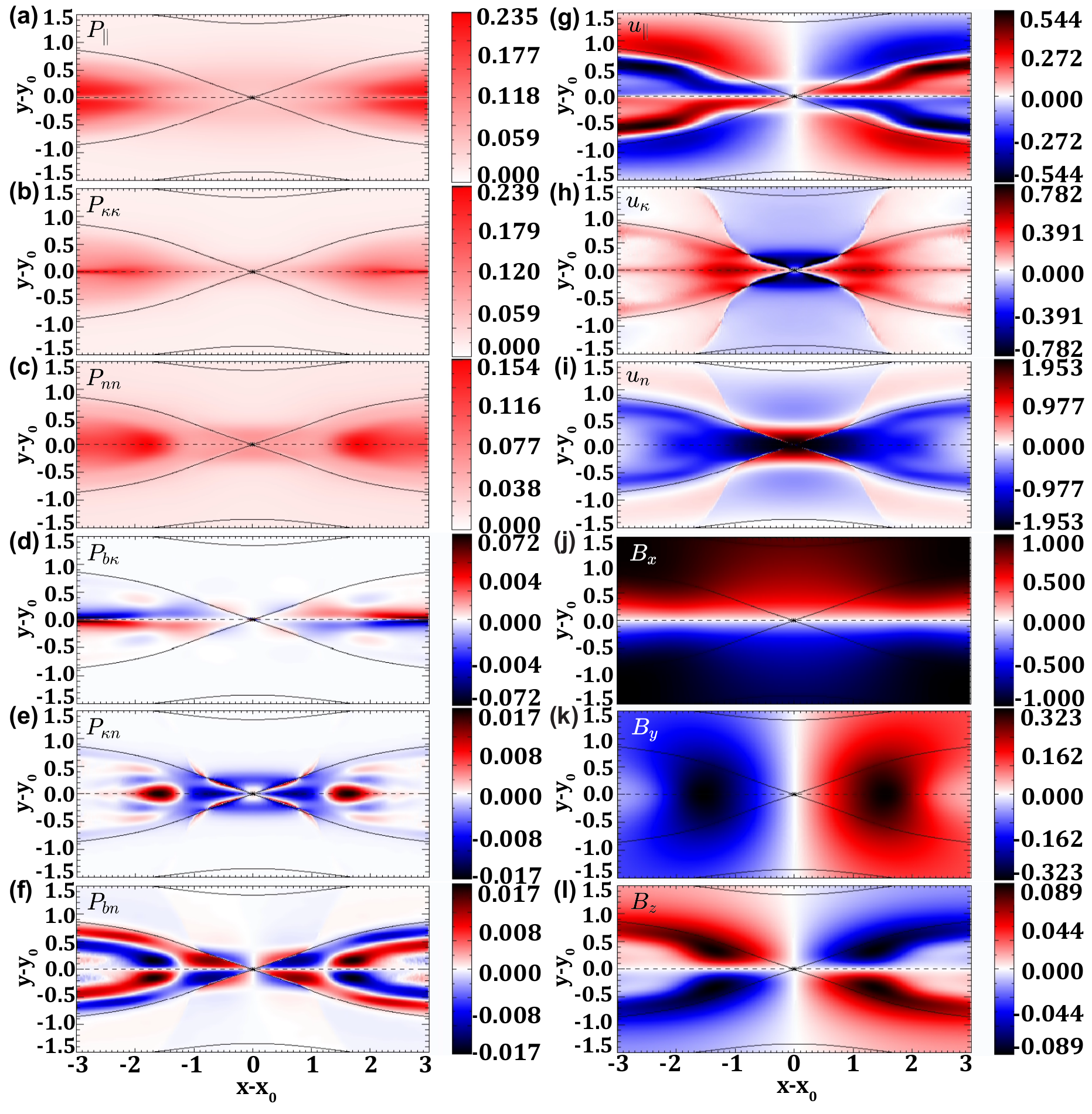}
\caption{\label{fig:2dprofiles_fieldaligned} For the same data as in Fig.~\ref{fig:2dprofiles}, 2D profiles of the (a)-(f) six independent elements of the electron pressure tensor ${\bf P}$ and (g)-(i) three components of bulk electron velocity ${\bf u}$ in field-aligned coordinates. The magnetic field ${\bf B}$ is plotted again in (j)-(l) for convenience. Representative magnetic field projections in the $xy$ plane are in black.}
\end{figure}

Figure~\ref{fig:2dprofiles_fieldaligned} displays 2D profiles of the plasma quantities, analogous to those in panels Fig.~\ref{fig:2dprofiles}(a)-(l) except the electron pressure tensor ${\bf P}$ [panels (a)-(f)] and the electron bulk flow velocity ${\bf u}$ [panels (g)-(i)] are in field-aligned coordinates with subscripts $b, \kappa,$ and $n$.  The magnetic field components [panels (j)-(l)] are repeated from Fig.~\ref{fig:2dprofiles} for convenience. 

The curvature direction is mostly in the $\pm x$ direction along $y = y_0$ and the $\pm y$ direction along $x = x_0$.  There is an abrupt change in the direction of $\boldsymbol{\hat{\kappa}}$ in the upstream region where the direction of the curvature of the magnetic field lines flips, which is particularly evident in Figs.~\ref{fig:2dprofiles_fieldaligned}(h) and (i).  The magnetic field curvature $\kappa$ is highest in the exhaust region near $y = y_0$ due to the curvature of the reconnected field lines. The binormal direction is mostly in the $\pm z$ direction except in the region of the out-of-plane Hall magnetic field where ${\bf \hat{n}}$ develops an in-plane component because the Hall effect bends the reconnecting magnetic field out of the reconnection plane within the diffusion region \cite{Sonnerup79,Mandt94}, as seen in Fig.~\ref{fig:2dprofiles_fieldaligned}(l). (There would be torsion in the upstream region if there was an initial out-of-plane guide magnetic field.)  There is also a strong torsion where $\boldsymbol{\hat{\kappa}}$ abruptly switches signs in the upstream region, since this causes an abrupt change in the ${\bf \hat{n}}$ direction and $\tau = - \boldsymbol{\hat{\kappa}} \cdot \nabla_\| {\bf \hat{n}}$.  However, the contribution to the pressure-strain interaction associated with this strong $\tau$ in the upstream region is weak. The torsion due to the Hall magnetic field has the same sign in the first and third quadrants relative to the X-line, and the opposite sign in the second and fourth quadrants. In our simulations, it is negative in the first quadrant close to the X-line $(x-x_0 < 0.2)$, then becomes positive from $0.2 < x-x_0 \lesssim 0.6$, and negative again further out in the EDR.

For the spatial structure of the plasma properties in field-aligned coordinates, the diagonal elements of the electron pressure tensor broadly have similarities in the Cartesian and field-aligned coordinate system, but the off-diagonal elements look very different. The bulk flow profiles are largely as expected. The parallel bulk flow $u_{\|}$ [Fig.~\ref{fig:2dprofiles_fieldaligned}(g)] is field aligned or anti-field aligned in the exhaust between the separatrices due to the change of direction of the magnetic field [Fig.~\ref{fig:2dprofiles_fieldaligned}(j)], with a similar pattern with reversed polarity due to the inflow outside the separatrix. This gives rise to an overall octupolar structure around the X-line. The flow in the direction of the curvature $u_{\kappa}$ is negative in the upstream region where the flow opposes the magnetic field curvature and positive in the exhaust where it is along the curvature [Fig.~\ref{fig:2dprofiles_fieldaligned}(h)].  The out-of-plane velocity $u_{z}$ [Fig.~\ref{fig:2dprofiles}(i)] is slightly negative in the IDR and mostly positive in the EDR, so the sign flips within those regions in the binormal component of the velocity $u_{n}$ [Fig.~\ref{fig:2dprofiles_fieldaligned}(i)] are due to the changing direction of ${\bf \hat{n}}$.

\begin{figure}
\includegraphics[width=5.in]{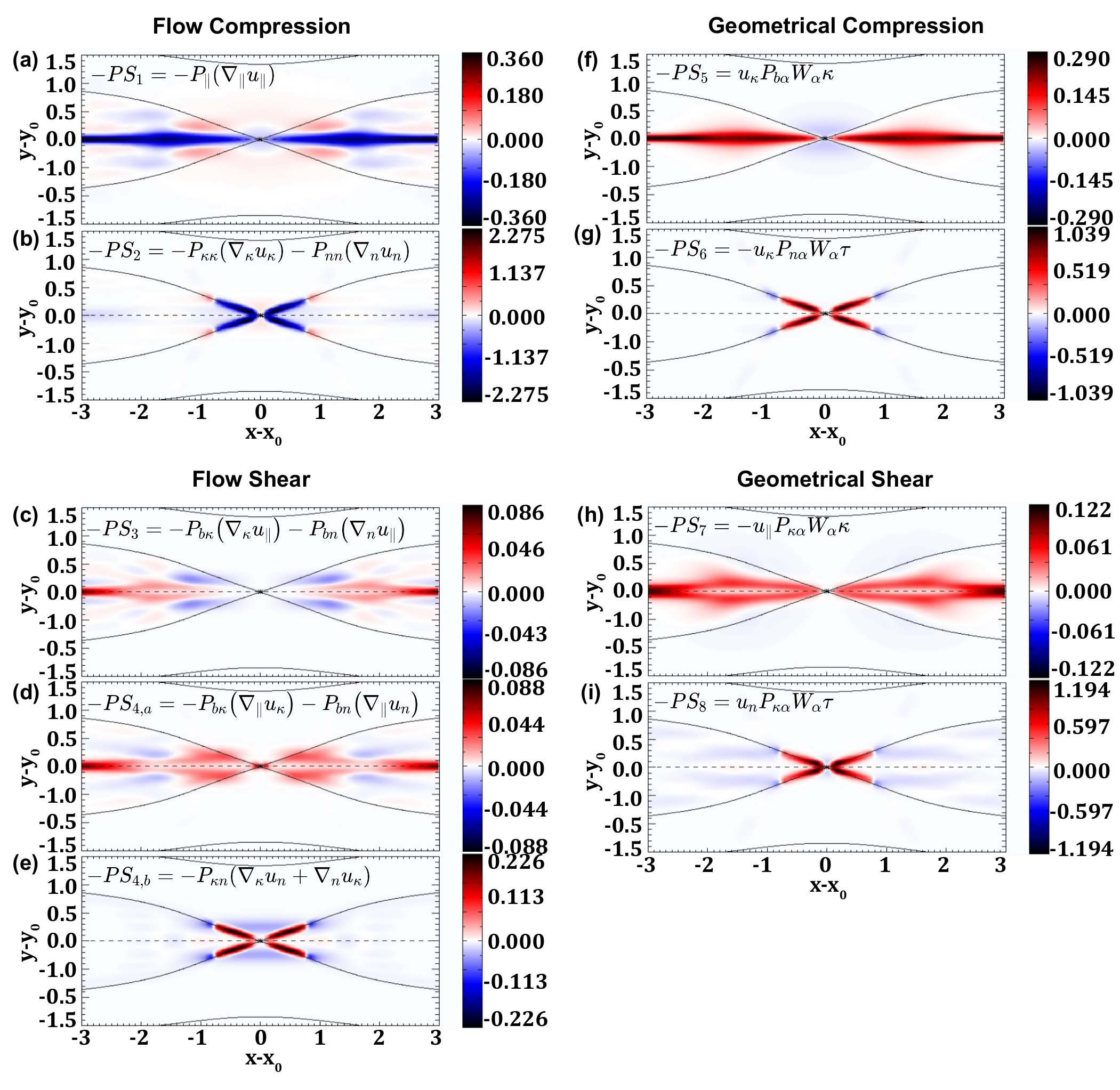}
\caption{\label{fig:9TermDecomp} Decomposition of the pressure-strain interaction $-({\bf P} \cdot \boldsymbol{\nabla}) \cdot {\bf u}$ for electrons in field-aligned coordinates classified according to their physical causes. Bulk flow compression in the (a) parallel $-PS_1$ and (b) perpendicular $-PS_2$ directions.  Bulk flow shear for (c) parallel flow varying in the perpendicular direction $-PS_3$, (d) perpendicular flow varying in the perpendicular direction $-PS_{4,a}$, and (e) perpendicular flow varying in the parallel direction $-PS_{4,b}$.  Geometrical compression terms due to (f) perpendicular flow $-PS_5$ and (g) torsion $-PS_6$.  Geometrical shear terms due to (h) parallel flow $-PS_7$ and (i) torsion $-PS_8$.}
\end{figure}

\subsection{Decomposition of Pressure-Strain Interaction in Field-Aligned Coordinates}
\label{sec:analysis}

We plot the contributions to the pressure-strain interaction in field-aligned coordinates in Fig.~\ref{fig:9TermDecomp}. We briefly discuss each term in turn and relate their most significant structures to the physics of the reconnection process as an example of the utility of the method.

We start with bulk flow compression.  Parallel flow compression [$-PS_1$, panel (a)], is largest along $y = y_0$.  There, electrons are accelerated in the exhaust and obtain a component of flow parallel to the reconnected magnetic field that is positive in the first and third quadrants and negative in the second and fourth. At $y = y_0$, the magnetic field is mostly in the $y$ direction, so this abrupt change in $u_\|$ getting faster in the direction of the flow is expansion, so this is associated with a negative contribution to the pressure-strain interaction. Perpendicular flow compression [$-PS_2$, panel (b)] is extremely large along a portion of the separatrix within a localized region near the X-line.  This arises because the magnetic field lines are strongly kinked at the separatrices, accelerating electrons into the outflow jet normal to the magnetic field.  This expansion is associated with a negative contribution to the pressure-strain interaction. 

Next, we treat the bulk flow shear.  The shear of the parallel flow in the perpendicular directions [$-PS_3$, panel (c)] is relatively weak within the EDR.  It is most significant downstream of the exhaust at $y-y_0 \simeq 0$, where the jet enters the larger magnetic island. Within the EDR but near the downstream edge, there is parallel flow in the $\pm y$ direction, that speeds up with distance away from the X-line in the $\boldsymbol{{\hat \kappa}} \simeq \pm {\bf \hat{x}}$ direction. Due to the structure of $P_{b\kappa}$ [Fig.~\ref{fig:2dprofiles_fieldaligned}(d)] this term contributes to relatively weak positive pressure-strain interaction. Within the EDR but closer to the separatrices in the exhaust, the same parallel flow decreases in the $\boldsymbol{\hat{\kappa}}$ direction, which is expansion. Since $P_{b\kappa}$ has the same sign in this region as the downstream EDR edge, this contributes to very weak negative pressure-strain interaction. Parallel shear of the perpendicular flow [$-PS_{4,a}$, panel (d)] is strongest in the EDR region in the exhausts just inside the separatrices. This is caused by $u_n$, which has a component in the outflow direction because of the bending of the reconnected field by the Hall effect, that changes along the magnetic field in a region of non-zero $P_{bn}$.  This contribution to pressure-strain interaction is rather weak; there is stronger heating downstream of the EDR near $y = y_0$. Perpendicular shear of the perpendicular flow [$-PS_{4,b}$, panel (e)] is strongest in the separatrix region near the X-line, where the terms due to the outflow mostly in $u_\kappa$ and the current sheet flow mostly in $u_n$ contribute a comparable amount. This leads to a contribution towards positive pressure-strain interaction. Also, at the upstream edge of the EDR, $u_n$ is predominantly in the out-of-plane (${\bf \hat{z}}$) direction and varies in the inflow ($\pm {\bf \hat{y}}$) direction, which is opposite to the curvature direction in this region.  This conspires with the negative $P_{\kappa n}$ to give a negative contribution to the pressure-strain interaction.

Turning to the geometrical terms, perpendicular geometrical compression [$-PS_5$, panel (f)] has its dominant signal near $y = y_0$, where the strongly curved field lines drive the outflow jet in the direction of the magnetic curvature, {\it i.e.,} the bulk flow in the exhaust has a strong positive perpendicular component $u_\kappa$ [Fig.~\ref{fig:2dprofiles_fieldaligned}(h)].
Predominantly due to the diagonal element $P_\|$ [see Fig.~1(e) in Paper II], this gives a positive contribution to the pressure-strain interaction. As emphasized in Paper II, no contribution to the pressure-strain interaction is due to direct heating by the magnetic field; rather the flows relative to the curve of the magnetic field line in this case are convergent, leading to geometrical compression.  In the regions upstream of the X-line within the EDR, the bulk flow is in the opposite direction to the curvature, so it contributes towards negative pressure-strain interaction but much more weakly than in the exhaust. Torsional geometrical compression [$-PS_6$, panel (g)] has a significant contribution towards positive pressure-strain interaction at the separatrix near the X-line, which is due to the torsion generated by the in-plane magnetic field lines being dragged out of the page due to the Hall effect. It is strongest at the separatrices where the inflow is initially accelerated into the outflow, generating a positive $u_\kappa$ [see Fig.~\ref{fig:2dprofiles_fieldaligned}(h)]. The $P_{nn}$ term leads to the strongest contribution to positive pressure-strain interaction. Next is parallel geometrical shear [$-PS_7$, panel (h)]. It is strongest downstream of the EDR, but also has positive signal inside the EDR in the outflow edges confined within the separatrices and away from $y = y_0$. This occurs because there is a diagonal pressure tensor element $P_{\kappa \kappa}$ coinciding with a parallel velocity $u_\|$ in the curved magnetic fields of the exhaust, and this leads to a contribution towards positive pressure-strain interaction [see Fig.~1(g) of Paper II]. Finally, torsional geometrical shear [$-PS_8$, panel (i)] describes a positive contribution to pressure-strain interaction localized to the separatrix near the X-line, as with the other torsional geometrical term $-PS_6$ [panel (g)]. This occurs because there is a flow due the projection of the out-of-plane electron flow in the binormal direction, which conspires with $P_{\kappa \kappa}$ to contribute to give a positive contribution to the pressure-strain interaction.

\subsection{Largest Contributions to the Pressure-Strain Interaction in Field-Aligned Coordinates}
\label{sec:contributions}

Having treated the terms individually, now we discuss the terms that dominate the regions where we see the most important features of the pressure-strain interaction.  Consider first the region immediately surrounding the X-line, where Fig.~\ref{fig:2TermDecomp}(a) shows that there is a local negative contribution at the X-line that extends in the outflow direction.  The physical cause of this feature is that the outflow $u_{e,out}$ accelerates from rest to the peak outflow speed over a distance $L_e$ in the outflow direction.  Thus, the perpendicular flow $u_\kappa$ [see Fig.~\ref{fig:2dprofiles_fieldaligned}(h)] is expanding in the direction it is pointing, which is associated with cooling from $-PS_2$. 

We perform a scaling analysis of this term to estimate its contribution quantitatively.  The $\kappa$ term of $-PS_2$, namely $-P_{\kappa \kappa} (\nabla_\kappa u_\kappa)$, scales like $-PS_2 \sim -P_{\kappa\kappa} (u_{e,out}/L_e)$.  For $P_{\kappa\kappa}$, this is $P_{xx}$ in the EDR, and for scaling purposes we take this to be the upstream electron pressure $P_{e,up}$ [which is justified by Fig.~\ref{fig:2dprofiles_fieldaligned}(b)].  For the outflow speed, we expect that during steady state reconnection, it scales as the electron Alfv\'en speed $c_{A,eup}$ based on the magnetic field strength $B_{e,up}$ upstream of the EDR.  For the length of the EDR in the outflow direction, it scales as approximately $5 \ d_e$, where $d_e$ is the electron inertial scale.  Putting these together and using $c_{Ae,up}/d_e = \Omega_{ce,up}$, the electron cyclotron frequency based on $B_{e,up}$, we get $-PS_2 \sim - 0.2 P_{e,up} \Omega_{ce}$.  We expect this to hold in the steady state, but our simulation data is taken during the onset phase instead.  To test the scaling, we therefore use the empirically measured $u_{e,out} \simeq u_\kappa \simeq 0.8$ from Fig.~\ref{fig:2dprofiles_fieldaligned}(h).  With simulation parameters $P_{e,up} = n_e T_{e,up} = 0.017$ and $L_e = 5 \ d_e = 2.2$, we get $-PS_2 \simeq -0.006$. This is in reasonable agreement with the simulated value of 0.009 for $-({\bf P} \cdot \boldsymbol{\nabla}) \cdot {\bf u}$ at the X-line in Fig.~\ref{fig:2TermDecomp}(a). 

Further away from the X-line in the outflow direction along $y-y_0 = 0$ line, the pressure-strain interaction becomes strongly positive.  In field-aligned coordinates, this happens because the outflow jet has a significant component parallel to the curvature direction leading to perpendicular geometrical compression $-PS_5$.  The associated contribution scales as $-PS_5  \simeq u_\kappa P_\| \kappa \sim u_{e,out} P_{e,up} / d_e \sim P_{e,up} \Omega_{ce,up}$. The gradient scale in this case is $d_e$, the thickness of the EDR in the $y$ direction, since that is the gradient that comes into the calculation of the curvature $\kappa$, and we similarly take $P_\| \sim P_{e,up}$. This heating rate is 5 times higher than the cooling rate near the X-line discussed in the previous paragraph. For the simulations, this gives $-PS_5 \simeq 0.03$. This is in good agreement with the measured value of $-({\bf P} \cdot \boldsymbol{\nabla}) \cdot {\bf u}$ in this region of 0.05, as seen in  Fig.~\ref{fig:2TermDecomp}(a).

\begin{figure}
\includegraphics[width=3.4in]{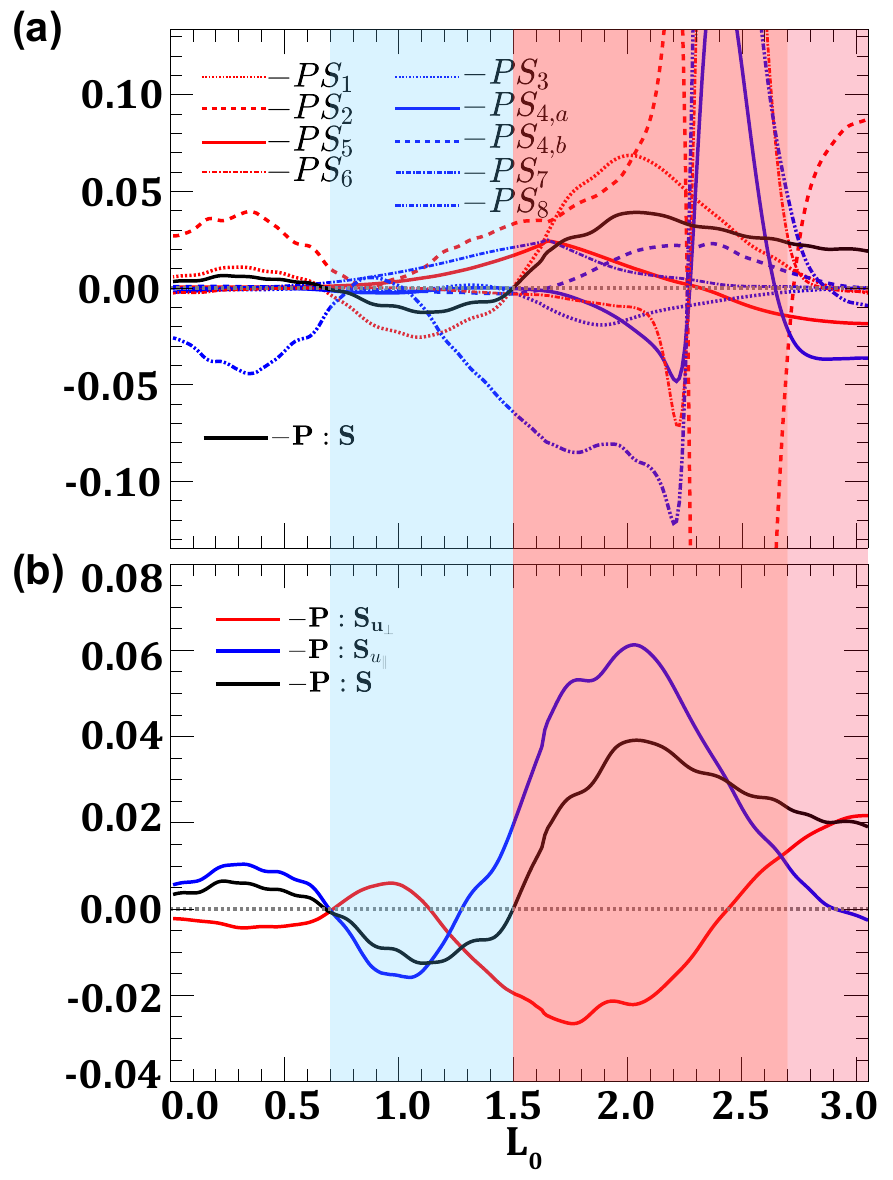}
\caption{\label{fig:9TermDecomp1Dcuts} Pressure-strain interaction along the 1D cut shown as the dotted path in Fig.~\ref{fig:2TermDecomp}(a).  $L_0$ is the distance from the left along the dotted path.  In both panels, the pressure-strain interaction $-{\bf P}:{\bf S}$ is in black.  (a) Contribution due to each $-PS_j$ term, with compression terms in red and shear terms in blue. (b) The pressure-strain interaction contribution $-{\bf P}:{\bf S}_{u_\|}$ dependent on $u_\|$   (blue) and the pressure-strain interaction contribution $-{\bf P}:{\bf S}_{{\bf u}_\perp}$ dependent on $u_\kappa$ and $u_n$ (red).}
\end{figure}

The other main regions of non-zero pressure strain are the upstream edges of the EDR and the region of cooling at $1.5 < |x-x_0| < 2.5$ and $|y-y_0| \simeq 0.35$.  To investigate the dominant terms, we take a cut along the dotted path in Fig.~\ref{fig:2TermDecomp}(a), the same path used to make Fig.~\ref{fig:1dcuts}.  The result is shown in Fig.~\ref{fig:9TermDecomp1Dcuts}, with panel (a) showing the pressure-strain interaction in black, and the nine terms $-PS_j$ (with separate lines for $-PS_{4,a}$ and $-PS_{4,b}$) in red (compression terms) and blue (shear terms) lines.  Panel (b) again shows the pressure-strain interaction in black, but here the sum of all of the terms dependent on $u_\|$ are given in blue, while the sum of all the terms dependent on $u_\kappa$ and $u_n$ are given in red.  We consider three different regions on this plot: immediately upstream of the X-line ($2.7 < L_0 < 3.0$, shaded pink), between this region and the separatrix in the upstream region ($1.5 < L_0 < 2.7$, shaded red), and the region of negative pressure-strain interaction ($0.7 < L_0 < 1.5$, shaded blue).  

Immediately upstream of the X-line (or more appropriately for the present purposes, the stagnation point), the contribution to positive pressure-strain interaction is caused by electrons slowing down from their inflow speed to a speed of 0 at the stagnation point, which is a convergent (compressional) perpendicular flow. Thus, $-PS_2$ is the dominant contributor to the observed pressure-strain interaction, appearing as the red region near $x = x_0$ in Fig.~\ref{fig:9TermDecomp}(b), and showing up as the dominant contribution in the range $2.7 < L_0 < 3.0$ as the dashed red line in Fig.~\ref{fig:9TermDecomp1Dcuts}(a). We find that the dominant contribution to $-PS_2$ is $-P_{nn} (\nabla_n u_n)$ (not shown). To estimate this term with a scaling analysis, we again take $P_{nn} \sim P_{e,up}$, so $-PS_2 \sim P_{e,up} u_n/L_n$, where $u_n$ is the characteristic binormal speed and $L_n$ is the gradient scale in the $n$ direction.  In this region, the magnetic fields bend out of the plane because of the Hall effect \cite{Sonnerup79,Mandt94}, so the ${\bf \hat{n}}$ direction has components in the ${\bf \hat{z}}$ and the $\pm {\bf \hat{y}}$ directions.  Since the Hall $B_z$ field strength scales as \cite{Mandt94} the reconnecting field strength $B_x$, the angle $\theta$ that ${\bf \hat{n}}$ makes with the reconnection plane scales as $45^\circ$.  Then, we estimate $u_n$ knowing that the dominant bulk flow in this region is the current sheet flow in the ${\bf \hat{z}}$ direction. A scaling analysis using Amp\`ere's law gives (in cgs units) $u_z \sim c (\partial B_x/\partial y) / 4\pi e n_e \sim c B_{e,up} / 4\pi e n_e d_e \sim c_{Ae,up}$. Projecting this into the $n$ direction gives $u_n \sim u_z \cos\theta \sim c_{Ae,up} \cos\theta$. Similarly, we estimate $L_n$ by noting the primary direction of variation is $y$.  This gives $L_n \sim L_y \cos\theta \sim d_e \cos\theta$. Putting it all together, we get $-PS_2 \sim P_{e,up} \Omega_{ce,up} \cos^2 \theta$. With $\theta \sim 45^\circ$, this term scales as approximately half the value of $-PS_5$ in the exhaust. For the simulations, we use the empirically determined $u_z \simeq 2$ to get a scaling prediction of $-PS_2 \sim P_{e,up} (u_z / d_e) \cos^2\theta \simeq 0.04$. This is a factor of two lower compared to the heating rate of 0.09 given by the dashed red line at $L_0 = 3.0$ in Fig.~\ref{fig:9TermDecomp1Dcuts}(a), which reflects the significant assumptions made in our estimates.

In the region of positive pressure-strain interaction $1.5 < L_0 < 2.7$ leading up to the separatrix, Fig.~\ref{fig:9TermDecomp1Dcuts}(a) reveals that a complicated mixture of terms play a role, with significant cancellation in parts.  Figure~\ref{fig:9TermDecomp1Dcuts}(b) makes an assessment of the contributions more transparent, showing that the terms associated with the parallel flow are the main contributors. In the region of interest, a positive pressure-strain interaction occurs due to $-PS_1$ because the parallel velocity of the inflowing electrons changes direction at the upstream edge of the EDR [Fig.~\ref{fig:2dprofiles_fieldaligned}(g)] which is a flow convergence and is therefore contributes to positive pressure-strain interaction. To quantify this with a scaling analysis, $-PS_1 \simeq P_\| \Delta u_{e,in \|} / L_\|$, where $\Delta u_{e,in \|}$ is the change in parallel inflow speed at the upstream edge of the EDR, and $L_\|$ is the length-scale over which $u_{e,in \|}$ changes directions.  It is difficult to estimate the change in flow and the gradient scale in terms of the upstream parameters, so we use values empirically determined from the simulations.  Using $L_\| \sim d_e \sim 0.45$ and $\Delta u_{e,in \|} \simeq 0.4$ from Fig.~\ref{fig:2dprofiles_fieldaligned}(g), and $P_\| \sim P_{e,up} \sim 0.017$, we get $-PS_1 \simeq 0.015$. This is in reasonable agreement with the values of $-PS_1$ in the region of interest in Fig.~\ref{fig:9TermDecomp1Dcuts}(a), where the dotted red curve varies from 0 to $\simeq 0.07$ with an average of $\simeq 0.03$. This is less than the heating rate due to $-PS_5$ in the exhaust, and we expect it would also be smaller than $-PS_5$ for a realistic system.

Finally, at the location where the pressure-strain interaction is negative ($0.7 < L_0 < 1.5$), we see from both panels of Fig.~\ref{fig:9TermDecomp1Dcuts} that parallel flow compression $-PS_1$ is the dominant term.  This is consistent with the hypothesis in Sec.~\ref{sec:contributions1} that this negative pressure-strain interaction is caused by the separatrix opening out while reconnection is getting faster during its onset phase. To estimate the amplitude via a scaling analysis, it is $-PS_1 \sim P_\| \Delta u_{e,out \|} / L_{out,\|}$, where $\Delta u_{e,out \|}$ is the parallel speed in the exhaust region at the location of interest and $L_{out,\|}$ is the length scale over which it changes, {\it i.e.,} the distance to the X-line.  This is again difficult to estimate in terms of the upstream parameters, but the empirical simulation results are $u_{e,out \|} \simeq -0.5$ and $L_{out,\|} \simeq 2$, so $-PS_1 \simeq -0.004$.  Fig.~\ref{fig:9TermDecomp1Dcuts}(a) gives a value of -0.025, about a factor of 6 higher, which again reflects the roughness of the estimation.

\section{Discussion and Conclusions}
\label{sec:discuss}

This study concerns using the pressure-strain interaction to study the rate of conversion between bulk flow and thermal energy density during magnetic reconnection. Using two-dimensional particle-in-cell simulations of anti-parallel symmetric reconnection and the analyses in Paper I and Paper II, we calculate decompositions of the pressure-strain interaction in Cartesian and magnetic field-aligned coordinates in and around the EDR.  

One purpose of this study is to demonstrate how to use the results of Paper I and Paper II to analyze a physical system. In so doing, we plot the decomposition of the pressure-strain interaction in terms of the pressure dilatation and ${\rm Pi-D}$ (compressible and incompressible contributions, respectively), and compare it to the decomposition from Paper I with ${\rm PDU}$ and ${\rm Pi-D}_{{\rm shear}}$ terms (flow convergence/divergence and flow shear, respectively).  We find their structure is noticeably different. Both decompositions have their merit in isolating particular physical effects.  For the present study of reconnection, we find that a number of features of the most prominent contributions to the pressure-strain interaction are better isolated by employing ${\rm PDU}$ and ${\rm Pi-D}_{{\rm shear}}$, and significant insights are gained by using the two decompositions in tandem. We similarly calculate the decomposition of pressure-strain interaction in magnetic field-aligned coordinates. As desired, this decomposition facilitates a physical interpretation of the mechanisms for heating relative to the ambient magnetic field, and allows for quantitative estimates of the energy density conversion rate from scaling analyses.

\begin{figure}
\includegraphics[width=5.2in]{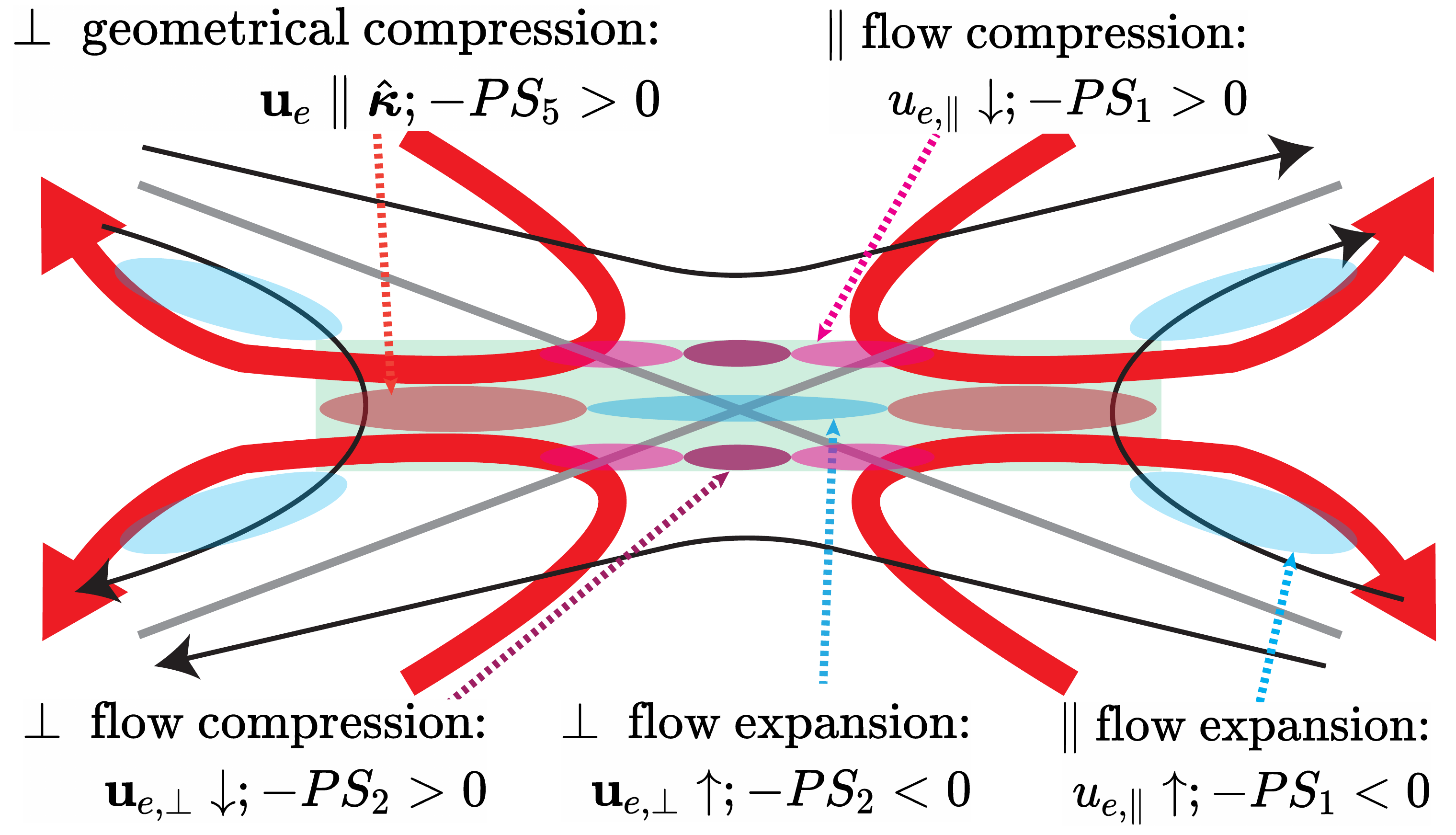}
\caption{\label{fig:2dreconsketch} Sketch of the physical mechanisms contributing to the pressure-strain interaction in a magnetic reconnection region electron diffusion region (EDR) during the reconnection onset phase.  In-plane projections of the magnetic field ${\bf B}$ are in black and gray, and the in-plane electron bulk flow ${\bf u}_e$ is in red. The green rectangle denotes the EDR. The ellipses in the red color palette denote regions of positive pressure-strain interaction (a contribution to heating), and the blue ellipses denote negative pressure-strain interaction (a contribution to cooling). The colored dashed arrows illustrate the physical mechanism causing the non-zero pressure-strain interaction in each location.}
\end{figure}

A second purpose of this study is to better understand the conversion of energy between bulk flow and thermal energy density during magnetic reconnection.  The result of this analysis is summarized by a map in   Fig.~\ref{fig:2dreconsketch} of where the different effects are most important near the EDR in our simulations during the onset phase of reconnection. It contains a sketch of a region around a reconnection X-line, with projections of the magnetic field in the reconnection plane in black, electron flow lines in red, and the EDR shaded green.  Ellipses denote regions for which the pressure-strain interaction is most appreciable, with colors in the red color palette denoting a contribution to positive pressure-strain interaction and blue denoting negative pressure-strain interaction.  Each set of regions has an arrow pointing to it describing the physical mechanism causing the positive or negative pressure-strain interaction as a result of the analysis in Sec.~\ref{sec:contributions}:
\begin{enumerate}
\item $-PS_2$ causes positive pressure-strain interaction at the upstream edges of the EDR above and below the X-line due to perpendicular compression as the electron inflow slows down.
\item $-PS_1$ causes positive pressure-strain interaction at the upstream edge of the EDR out to the separatrices due to parallel compression as the inflow of electrons slow down as they approach the EDR.
\item $-PS_2$ causes negative pressure-strain interaction at and in the near vicinity of the X-line because electrons experience expansion as they are accelerated in the outflow direction.
\item $-PS_5$ causes positive pressure-strain interaction at the downstream edge of the EDR due to perpendicular geometrical compression since the outflow has a component in the direction of the magnetic curvature. 
\item $-PS_1$ causes negative pressure-strain interaction in the downstream region due to expansion of the parallel flow, which is presumably associated the outflow jets being redirected in the vertical direction and speeding up while the separatrix opens out during the onset phase.
\end{enumerate}
It bears repeating that, in isolation, positive and negative pressure-strain interaction would be associated with a local increase and decrease in thermal energy density, respectively, {\it i.e.,} heating and cooling, but there are other terms that can locally change the thermal energy density so one cannot conclude there is local heating or cooling just from the sign of the pressure-strain interaction.  Also, as a reminder, the present simulations were carried out with relatively high electron mass; we expect the structures would look qualitatively similar for a more realistic mass ratio, but likely with sharper features and higher amplitudes. This should be tested in future work.

To apply these results to steady-state reconnection, we expect that mechanisms \#1-4 carry over relatively unchanged from the results here during reconnection onset.  However, the downstream negative pressure-strain interaction in \#5 is not likely to occur close to the EDR in steady-state reconnection. Instead, we expect it would occur in natural systems far downstream where the magnetic island grows. This is consistent with the absence of a coherent negative pressure-strain interaction in Fig.~2 of Ref.~\cite{Pezzi21} for steady-state reconnection. This raises the possibility that the presence of negative pressure-strain interaction near the downstream edge of the EDR could be used as a signal of reconnection being amidst its onset phase, but it would take further work beyond a single simulation to confirm or refute this possibility.

A key result of the present study is quantifying the expected scale of pressure-strain interaction in the EDR during magnetic reconnection. A simple scaling analysis reveals that the natural scale that describes heating via the pressure-strain interaction $-({\bf P} \cdot \boldsymbol{\nabla}) \cdot {\bf u}$ in an anti-parallel reconnection EDR is $\pm P_{e,up} c_{Ae,up} / (1-5 d_e)$ during the steady-state reconnection phase. In writing this, we use that the pressure in the EDR scales with the electron pressure $P_{e,up}$ upstream of the diffusion region, the bulk flow velocity scales with the electron Alfv\'en speed $c_{Ae,up}$ based on the magnetic field $B_{e,up}$ at the upstream edge of the EDR, and the gradient scale is either 1 or 5 $d_e$, depending on if the gradient is in the inflow or outflow direction, since $5 d_e$ is an expected relevant scale for the length of the EDR in the outflow direction.  This implies 
\begin{equation}
-{\bf P}:{\bf S}_e \sim \pm (0.2-1) P_{e,up} \Omega_{ce,up}, \label{eq:psscaling}
\end{equation}
where $\Omega_{ce,up} = c_{Ae,up}/d_e$ is the electron cyclotron frequency based on $B_{e,up}$.

This prediction should be useful for quantitative comparisons of the pressure-strain interaction during magnetic reconnection in space and the laboratory.  We treat a single case study as an example. The pressure-strain interaction was studied \cite{bandyopadhyay_energy_2021} during a magnetosheath reconnection event \cite{Wilder18}. Using plasma parameters of $B_{i,up} \simeq 40$~nT for the asymptotic (ion scale) reconnecting magnetic field, $n \simeq 10$~cm$^{-3}$ for the electron number density, and $T_{e,up} \simeq 70$~eV for the upstream electron temperature, and assuming the magnetic field at the electron layer scales as \cite{liu_FirstPrinciple_2022} $B_{e,up} \simeq (m_e/m_i)^{1/4} B_{i,up}$, we find $P_{e,up} \simeq 0.112$~nPa, $\Omega_{ce,up} \simeq 1.07 \times 10^3$~rad/s, and therefore $(0.2-1) P_{e,up} \Omega_{ce,up} \simeq 24-120$~nW/m$^3$. We expect a similar scaling relation as Eq.~(\ref{eq:psscaling}) to hold for ions except that the length scale in the outflow direction scales as $10 \ d_i$, so we expect
\begin{equation}
-{\bf P}:{\bf S}_i \sim \pm (0.1-1) P_{i,up} \Omega_{ci,up}.
\end{equation}
For the same event \cite{Wilder18}, $T_{i,up} \simeq 800$~eV and $n_i \simeq n_e$, so $P_{i,up} \simeq 1.28$~nPa and $\Omega_{ci,up} \simeq 3.8$~rad/s which implies $(0.1-1)P_{i,up} \Omega_{ci,up} \simeq (0.49-4.9)$~nW/m$^3$.  The measured pressure-strain interaction terms for electrons and ions peaked in the 30-40 nW/m$^3$ range and just above 2~nW/m$^3$, respectively \cite{bandyopadhyay_energy_2021}. The ratio of the electron heating rate to the ion heating rate scales as $(T_{e,up}/T_{i,up}) (B_{e,up}/B_{i,up}) (m_i/m_e) \sim (T_{e,up}/T_{i,up}) (m_i/m_e)^{3/4}$.
For the MMS event, this is $\simeq 25$, compared to the measured ratio of about 20. Thus, the present scaling predictions are in good agreement with the observations of this event, both for the absolute scale for electrons and ions, and for the ratio between the electron and ion heating rates.

The research presented here reveals some important insights. The pressure-strain interaction is independent of the coordinate system in use, whether in a Cartesian coordinate system or one in which the coordinate system is curved.  We have confirmed this for the simulations presented here (not shown). 

However, the contributions to the pressure-strain interaction from compression/expansion and bulk flow shear are strongly dependent on the coordinate system, as anticipated in Paper II.  Thus, the physical mechanism leading to the dominant pressure-strain interaction need not be the same in different coordinate systems. This is vividly seen for the example of magnetic reconnection treated here, with the plots shown in Figs.~\ref{fig:1dcuts} and \ref{fig:9TermDecomp1Dcuts} in Cartesian coordinates and field-aligned coordinates, respectively. We find in the region $0.7 < L_0 < 1.5$ that the negative pressure-strain interaction has contributions in Cartesian coordinates from both ${\rm PDU}$ and ${\rm Pi-D}_{{\rm shear}}$, with ${\rm PDU}$ being the dominant contributor; the mechanism in magnetic field-aligned coordinates is parallel flow expansion. This is the same physical mechanism, although it is not possible to identify from the Cartesian decomposition that the expansion is largely in the parallel direction. (This also illustrates one benefit of employing the magnetic field-aligned coordinate system.) However, for the positive pressure-strain interaction contribution ($1.5 < L_0 < 2.7$), it was found in Cartesian coordinates that the dominant contribution is ${\rm Pi-D}_{{\rm shear}}$, {\it i.e.,} bulk flow shear, while the magnetic field-aligned coordinate result is parallel flow compression. This exposes a potential pitfall in analyzing decompositions of the pressure-strain interaction contributions: the physical mechanisms in a Cartesian coordinate system may be different in a magnetic field-aligned coordinate system.

A second pitfall could arise in determining the dominant contribution to the pressure-strain interaction.  If one wants to find the term in the decomposition that leads to the highest values, one might find the $-PS_j$ terms that are the largest and identify them as the most important.  If we were to do that in the present study of reconnection, we would find that $-PS_2$ [Fig.~\ref{fig:9TermDecomp}(b)] is the largest due to its contribution at the separatrix region near the X-line.  However, in this region the perpendicular flow shear [$-PS_{4,b}$, Fig.~\ref{fig:9TermDecomp}(e)], torsional geometrical compression [$-PS_6$, Figs.~\ref{fig:9TermDecomp}(g)], and torsional geometrical shear [$-PS_8$, Figs.~\ref{fig:9TermDecomp}(i)] are also important. By comparing amplitudes individually and with the pressure-strain interaction [$-({\bf P} \cdot \boldsymbol{\nabla}) \cdot {\bf u}$, panel (a)], we see that perpendicular flow compression is more than 40 times larger the maximum value of the pressure-strain interaction.  Moreover, the pressure-strain interaction does not display a feature at the separatrices near the X-line where these signals are strongest.  This implies that the four terms cancel each other nearly completely in this region, leaving relatively weak pressure-strain interaction at the EDR separatrix.  Thus, the decomposition of pressure-strain interaction in may lead to individual terms that are much larger than the total.  Instead, finding the dominant term should be carried out by finding the pressure-strain interaction first, then finding which terms contribute most strongly in the region of interest.

There is a physical reason that the terms in the decomposition in field-aligned coordinates can be significantly larger than the pressure-strain interaction itself, with significant cancellation between terms.  Field-aligned coordinates follow magnetic field lines. At the separatrices the flow lines are strongly kinked. The strong kink leads to a huge velocity shear, which contributes to positive pressure-strain interaction.  But the plasma is strongly accelerated as well, leading to negative pressure-strain interaction through expansion.  For both, the gradient length scale is set by the scale over which the flow profile changes.  As seen in Fig.~\ref{fig:2dprofiles_fieldaligned}, the gradient in the flow can occur on scales far below the electron inertial scale.  In our simulations, the gradient scale could be as low as the electron Debye scale or the grid scale, both less than 0.02 in normalized units.  This is about 20-40 times smaller than $d_e$, which explains why the terms mentioned in the previous paragraph can be 20-40 times the pressure-strain interaction in total.  We do not attempt in this study to ascertain whether the Debye scale or grid scale sets the gradient scale and instead leave that for future work; it can be easily studied by varying the two scales relative to each other.

Finally, it bears repeating that the pressure-strain interaction is a local measure of the rate of energy conversion between bulk flow and thermal energy density, but it is not the full local measure of heating or cooling.  It remains true that in an infinite or closed and isolated system, the global energy conversion is governed by the pressure-strain interaction \cite{Yang17,Yang_2022_ApJ}, but thermal energy flux, heat flux, and enthalpy flux can also change the local thermal energy density \cite{Du20,song_forcebalance_2020}. Moreover, there are other metrics for the rate of other kinds of energy conversion in plasma processes, such as ${\bf J} \cdot {\bf E}$ \cite{Zenitani11} and its kinetic counterpart, the field-particle correlation \cite{Klein16}, which have received a lot of attention \cite{Burch16b,Wilder18,Chen19,Afshari21} because they describe the volumetric rate of conversion between the bulk flow energy and the energy in the electromagnetic fields. There have been studies comparing these other metrics with the pressure-strain interaction \cite{Pezzi21}; it would be interesting to revisit such studies in light of the results of the present series of papers.  A second important point is that thermal energy is merely one form of energy in a plasma not in local thermodynamic equilibrium.  The pressure-strain interaction does not provide information about energy conversion into other forms of energy, such as $\int (1/2) m v_x^{\prime} v_y^\prime f d^3v$ or higher order moments.  A measure of the energy conversion associated with moments of the phase space density other than the thermal energy density is treated in a separate study \cite{Cassak_FirstLaw_2022}.

\begin{acknowledgments}
We acknowledge beneficial conversations with Yan Yang.  We gratefully acknowledge support from NSF Grant PHY-1804428, DOE grant DE-SC0020294, and NASA grant 80NSSC19M0146.  This research uses resources of the National Energy Research Scientific Computing Center (NERSC), a DOE Office of Science User Facility supported by the Office of Science of the US Department of Energy under Contract no.~DE-AC02-05CH11231. Simulation data used in this manuscript are available on Zenodo \cite{Barbhuiya22Zenodo}.
\end{acknowledgments}


\providecommand{\noopsort}[1]{}\providecommand{\singleletter}[1]{#1}%

\end{document}